# A High Efficiency Aluminum-Ion Battery Using an $AlCl_3$-Urea Ionic Liquid Analogue Electrolyte

Michael Angell[1], Chun-Jern Pan[1,2], Youmin Rong[1], Chunze Yuan[1], Meng-Chang Lin[3], BingJoe Hwang[2], and Hongjie Dai[1*]

[1]Department of Chemistry, Stanford University, Stanford, California 94305, USA.

[2]Department of Chemical Engineering, National Taiwan University of Science and Technology, Taipei 10607, Taiwan.

[3]College of Electrical Engineering and Automation, Shandong University of Science and Technology, Qingdao 266590, P. R. China.

* Correspondence to hdai1@stanford.edu

## Abstract

In recent years, impressive advances in harvesting renewable energy have led to a pressing demand for the complimentary energy storage technology. Here, a high coulombic efficiency (~ 99.7%) Al battery is developed using earth-abundant aluminum as the anode, graphite as the cathode, and a cheap ionic liquid analogue electrolyte made from a mixture of $AlCl_3$ and urea in 1.3 : 1 molar ratio. The battery displays discharge voltage plateaus around 1.9 V and 1.5 V (average discharge = 1.73 V) and yielded a specific cathode capacity of ~73 mAh $g^{-1}$ at a current density of 100 mA $g^{-1}$ (~ 1.4 C). High coulombic efficiency over a range of charge-discharge rates and stability over ~150-200 cycles was easily demonstrated. In-situ Raman spectroscopy clearly showed chloroaluminate anion intercalation/deintercalation of graphite in the cathode side during charge/discharge and suggested the formation of a stage 2 graphite intercalation compound when fully charged. Raman spectroscopy and nuclear magnetic resonance suggested the existence of $AlCl_4^-$, $Al_2Cl_7^-$ anions and $[AlCl_2^{\cdot}(urea)_n]^+$ cations in the urea/$AlCl_3$ electrolyte when an excess of $AlCl_3$ was present. Aluminum deposition therefore proceeded through two pathways, one involving $Al_2Cl_7^-$ anions and the other involving $[AlCl_2^{\cdot}(urea)_n]^+$ cations. This battery is a promising prospect for a future high performance, low cost energy storage device.

## Introduction

Cheap, high-rate (fast charge/discharge) rechargeable batteries with long cycle lives are urgently needed for grid-scale storage of renewable energy, as it is becoming increasingly important to replace fossil fuels (1). Lithium-ion batteries (LIBs) are expensive and have limited cycle lives, which makes them non-ideal for grid-scale energy storage. Furthermore, high-rate capability is necessary for use in the grid, under which conditions LIBs become increasingly unsafe due to the flammability of the electrolytes used. Batteries based on aluminum offer a viable alternative due to aluminum's three-electron redox properties (offers potential for high capacity batteries), stability in the metallic state, and very high natural abundance. Furthermore, the development of these batteries based on non-flammable electrolytes of low toxicity is critical for minimizing safety hazard and environmental impact. For this reason, ionic liquids (ILs) have been investigated for energy storage due to their low vapor pressure and wide electrochemical windows, unfortunately with the caveat of high cost in most cases. A new class of ionic liquids, referred to as ionic liquid analogues (ILAs) or so-called deep eutectic solvents, generally formed through a mixture of a strongly Lewis acidic metal halide and Lewis basic ligand, have gained significant attention due to their comparable electrochemical and physical properties with reduced cost and minimal environmental footprint (2). Abbott et al. first disclosed an ILA derived from the mixture of $AlCl_3$ and an oxygen donor amide ligand (urea or acetamide), in which ions were formed through the heterolytic cleavage of $AlCl_3$ (the $Al_2Cl_6$ unit) giving $AlCl_4^-$ anions and $[AlCl_2(\text{ligand})_n]^+$ cations, with the latter shown to be responsible for reductive aluminum deposition (3). Since then, numerous different Lewis basic ligands have been shown to form ILAs when mixed with $AlCl_3$, which are capable of effective aluminum deposition (4-6).

Recently, our group developed a secondary Al battery system based on the reversible deposition/stripping of aluminum at the Al anode and reversible intercalation/deintercalation of chloroaluminate anions at the graphite cathode in a non-flammable 1-ethyl-3-methylimidazolium chloroaluminate (EMIC-$AlCl_3$) ionic liquid electrolyte (7, 8). A ratio of $AlCl_3$/EMIC = 1.3 by mol was used such that $Al_2Cl_7^-$ was present in the (acidic) electrolyte to facilitate aluminum deposition (9). During charging, $Al_2Cl_7^-$ is reduced to deposit aluminum metal, and $AlCl_4^-$ ions intercalate in graphite to oxidize carbon. During discharge, this battery exhibited a cathode specific capacity of ~ 70 mAh $g^{-1}$ with a coulombic efficiency of 97-98%, and ultra-high charge/discharge rate (up to 5000 mA $g^{-1}$) for over 7000 cycles. However, room for improvement exists as the parameter space for the Al battery remains largely unexplored. The three-electron redox properties of aluminum allow a theoretical specific anode capacity of 2980 mAh/g, so there is potential for much higher overall capacity (and energy density) of the battery through the investigation of new cathode and electrolyte materials (10-13). Furthermore, while the 97-98% coulombic efficiency of this battery is higher than those of most aqueous battery systems, there is still significant room for improvement. State-of-the-art LIBs are capable of 99.98% CE (14, 15), a benchmark that should be met by alternative battery systems. Another consideration is our existing Al battery electrolyte uses 1-ethyl-3-methylimidazolium chloride (EMIC), which is relatively expensive. Immediately plausible new electrolytes for this system could include any which are capable of reversible aluminum deposition/dissolution. In this work, we investigate the performance of a rechargeable Al battery using an ILA electrolyte based on urea, a superior compound in terms of cost (~ 50 times cheaper than EMIC) and eco-friendliness.

## Results and Discussion

**Cyclic voltammetry and galvanostatic charge/discharge of Al battery.** The battery cathode was constructed using a graphite powder/polymer binder pasted onto a carbon fiber paper substrate, and the anode was free-standing, high-purity Al foil. $AlCl_3$/urea electrolyte was kept below 40 $^oC$ during mixing to avoid electrolyte decomposition. Residual HCl impurities were removed by adding Al foil under heat and vacuum, followed by the addition of ethylaluminum dichloride (see Methods). Fig. 1 shows the cyclic-voltammetry (CV) of the Al anode and the graphite cathode in the $AlCl_3$/urea (by mol) = 1.3 electrolyte, a ratio we found yielded the battery with the highest capacity with good cycling stability. We observed several graphite oxidation peaks in the 1.6-2.0 V (vs. Al) range, while another well-defined peak appeared at ~2.05V (Fig. 1a). These processes, as well as the corresponding reduction events on the negative sweep, were easily correlated with the galvanostatic charge-discharge curve (Fig. 1b) for a battery with ~ 5 mg $cm^{-2}$ loading of active graphitic material. The redox processes are largely reversible but somewhat kinetically hindered, showing relatively wide peaks (Fig. 1a), most likely as a result of the high viscosity of the electrolyte (3). The CV of the deposition/dissolution of the aluminum anode (Fig. 1c) was also quite reversible. Based on the aluminum stripping/dissolution reaction and chloroaluminate anion intercalation in graphite, battery mechanisms are suggested and illustrated schematically in Figure 1d.

Figure 2 shows galvanostatic charge-discharge data for the Al-graphite cell using the $AlCl_3$/urea ILA electrolyte. Initial cycling at 100mA $g^{-1}$ required ~ 5-10 cycles for stabilization of the capacity and coulombic efficiency (CE), suggesting side reactions during this time. The CE during first cycle was consistently around 90%, and then (during the first 5-10 cycles) increased above 100 % until a stable capacity was reached (at which point CE was stabilized at ~ 99.7%) (Fig. 2a). The phenomenon of CE > 100% is unknown to the EMIC- $AlCl_3$ electrolyte system (7) and therefore might involve side reactions with the cationic aluminum species or unbound urea, a topic requiring further investigation. The boxed region of Fig. 2a (enlarged in Fig. 2b.) demonstrates the capacity at varied charge-discharge rate using two different cutoff voltages (2.2, 2.15 V—chosen based on Fig. 1a CV results), after which cycling at 100 mA $g^{-1}$ was resumed until ~180 cycles. A slight decay in CE was observed over this time but it remained > 99%. The effects of rate on the galvanostatic charge-discharge curves are shown in Fig. 2c, and reasonable capacities of ~75 mAh $g^{-1}$, 73 mAh $g^{-1}$, and 64 mAh $g^{-1}$ were obtained at 50 mA $g^{-1}$ (0.67 C), 100 mA $g^{-1}$ (1.4 C) , and 200 mA $g^{-1}$ (3.1 C) current densities, respectively.

**In situ Raman Spectroscopy.** In-situ Raman scattering during charging/discharging experiments (see Methods) were performed to investigate the changes to the graphite structure during battery operation. Figure 3 displays in-situ Raman spectra recorded during charge/discharge at a rate of 50 mA $g^{-1}$ correlated to the respective regions of the galvanostatic charge/discharge curve. The data were recorded in the battery's 81$^{st}$ charge-discharge cycle, without observing obvious increase in the graphite defect-related D-band (SI Fig. 1), suggesting high graphite structural integrity through chloroaluminate intercalation/deintercalation cycles. Immediately upon beginning the lower plateau charging process, the G-band of pristine graphite (1584 $cm^{-1}$) split by ~20 $cm^{-1}$. This splitting resulted from positive charge density arranging itself on the boundary layers of the graphite during intercalation. Boundary layers adjacent to intercalant layers experienced more positive charge density, leading to a large blue shift in the $E_{2g}$ band for these layers, giving rise to two different $E_{2g}$ peaks overall, inner (i) and outer (b) (Fig. 3a inset, spectra in red) (16, 17). Based on the ratio of the intensities of these two peaks, the intercalation stage (n > 2) at that moment in time could be calculated based on the following equation:

$$\frac{I_i}{I_b} = \frac{\sigma_i}{\sigma_b}\frac{(n-2)}{2}$$

where $\sigma_i/\sigma_b$ is the ratio of Raman scattering cross sections, which was assumed to be unity (16). This initial splitting therefore indicated the formation of a dilute stage 4-5 intercalation compound, and as charging continued the two peaks steadily blue shifted with increasing potential/capacity of the battery. The $E_{2g(b)}$ band then underwent a small splitting (~3 $cm^{-1}$) at 1.94-1.99 V. At this point, the stage number (n) was calculated to be approximately 2.5. Shortly afterwards (at 2.03 V) the $E_{2g(i)}$ band disappeared completely. This was followed by the $E_{2g(b)}$ roughly doubling in intensity before it underwent another large splitting (1619-1632 $cm^{-1}$) at the beginning of the upper plateau (~2.097 V) (Fig. 3c, spectra in blue). At the fully charged state, only one high intensity peak at 1632 $cm^{-1}$ remained, suggesting the formation of a stage 1 or 2 GIC since neither $E_{2g(i)}$ and $E_{2g(b)}$ bands were present(16) (Fig. 3b, spectra in blue). A stage 2 GIC was assumed, based on the capacity of the Al battery.

The subsequent discharge process was reflective of the charge process, demonstrating reversibility. As the discharge of the upper plateau began (2.011 V), there was a slight red shift by 1 $cm^{-1}$. This band then split (~12 $cm^{-1}$) halfway through the upper plateau (1.97 V), with another peak reappearing at 1619 $cm^{-1}$. The 1631 $cm^{-1}$ peak proceeded to completely disappear, and the 1619 $cm^{-1}$ peak maximized at ~1.66 V, which signified the end of the upper plateau discharging process/deformation of the stage 2 GIC (see Fig. 3d, spectra in blue). Half-way through the lower voltage plateau (1.535 V) the second large splitting occurred and the original $E_{2g(i)}$ began to reappear with decreasing potential (Fig. 3c inset, spectra in red). Shortly after the reappearance of the $E_{2g(i)}$ mode, another splitting occurred at 1.525-1.535 V, small in magnitude (~5 $cm^{-1}$), as was seen during the charging process (Fig. 3d, spectra in green). This splitting likely corresponded to one of the several lower current redox events in this region demonstrated by CV (Fig 1a.). Of course, all bands red-shifted during discharge.

**Speciation in electrolyte by Raman Spectroscopy.** Next, we investigated the speciation in several $AlCl_3$/urea electrolytes. In the $AlCl_3$/urea = 1.0 ILA electrolyte, it was suggested (3) that aluminum deposition must have occurred from a cationic species of the form $[AlCl_2^{\cdot} (ligand)_n]^+$, since $Al_2Cl_7^-$ was not present and $AlCl_4^-$ cannot be reduced in the relevant voltage window. We performed Raman spectroscopy studies of five electrolytes with $AlCl_3$/urea in the range of 1.0 - 1.5 (Fig. 4a). Raman spectroscopy has previously been used to reveal the existence of chloroaluminate anions in both ILs (18-20) and ILAs (21, 22), with the Raman shifts appearing rather invariant in both ILs or ILAs with different cationic species. We observed characteristic Raman shifts of $AlCl_4^-$ (311 $cm^{-1}$) and $Al_2Cl_7^-$ (347 $cm^{-1}$) for $AlCl_3$/urea > 1.0 (Fig. 4a). For the $AlCl_3$/urea = 1.0 electrolyte, only the 347 $cm^{-1}$ peak ($AlCl_4^-$) was present, supporting the absence of $Al_2Cl_7^-$. When more $AlCl_3$ was added (increasing to 1.1, 1.3, 1.4, 1.5 ratios), the peak at 310 $cm^{-1}$ ($Al_2Cl_7^-$) systematically intensified relative to 347 $cm^{-1}$, suggesting the existence of $Al_2Cl_7^-$. Additionally, we observed less intense modes of $Al_2Cl_7^-$ that also increased with $AlCl_3$ content (Fig. 4b) (19).

Since $Al_2Cl_7^-$ exists in our $AlCl_3$/urea = 1.3 electrolyte used for the Al battery, aluminum deposition likely occurs through two pathways (3, 9):

$$\text{Anode reaction: } 4\ Al_2Cl_7^- + 3\ e^- \rightarrow 7\ AlCl_4^- + Al \qquad \textbf{(1)}$$

$$\text{Anode reaction: } 2\ [AlCl_2^{\cdot} (urea)_2]^+ + 3\ e^- \rightarrow AlCl_4^- + 4\ (urea) + Al \qquad \textbf{(2)}$$

where deposition through a cationic species would likely be dominant (eq. 2). During aluminum deposition, cationic species will migrate to the aluminum electrode, while anionic species would migrate to the graphite electrode. Furthermore, Al deposition from the cation (via eq. 2) generates free urea at the surface of the aluminum electrode, which would likely react with some amount of $Al_2Cl_7^-$. Eq. 2 assumes that only a four-coordinate cation exists, in which two urea molecules are bound to Al by the oxygen atom in urea (3). The tri-coordinate cation is unlikely due to the lack of inductive substituents on the nitrogen that might allow it to be bidentate, as is seen in the case of acetamide derivatives (21). The intercalation process at the graphite cathode remains the same as in the EMIC-$AlCl_3$ Al battery case despite the different aluminum deposition/stripping processes at the anode:

$$\text{Cathode reaction: } AlCl_4^- + C_x - e^- \rightarrow C_x^+[AlCl_4]^- \qquad \textbf{(3)}$$

where $x$ is the number of carbon atoms per intercalated anion ($x$ = 30 based on a capacity of 75 mAh $g^{-1}$ from 50 mA $g^{-1}$ galvanostatic discharging data). The energy densities calculated using Eq. (1) and Eq. (2) were 45 Wh $kg^{-1}$ and 76 Wh $kg^{-1}$, respectively. These values represent an upper limit to the energy density, as the calculation neglects the fraction of neutral species that would necessarily accompany the anionic and cationic species in this liquid, which is not 100% ionic.

**Analyzing relative concentrations of ionic species in electrolyte.** We analyzed the relative concentrations of ions in the electrolyte, namely $[Al_2Cl_7^-]/[AlCl_4^-]$ and $[AlCl_2\cdot(urea)_2]^+/[Al_2Cl_7^-]$ using the ratio of the intensities of the Raman peaks of $Al_2Cl_7^-$ and $AlCl_4^-$ in the electrolyte (Fig. 4a). The ratio of the Raman scattering cross sections of $Al_2Cl_7^-$ and $AlCl_4^-$ anions has been derived for the EMIC/$AlCl_3$ system (20), and we used this value to estimate $[Al_2Cl_7^-]/[AlCl_4^-] = 0.6$ and $[AlCl_2\cdot(urea)_2]^+/[Al_2Cl_7^-] = 2.6$ in the $AlCl_3$/urea = 1.3 electrolyte. This further suggests that for $AlCl_3$/urea = 1.3 the aluminum deposition would be dominated by the cationic species, which is present at 2.6 times the concentration of $[Al_2Cl_7^-]$. The energy density of the real system based on only electrochemically active materials would therefore be closer to 76 Wh $kg^{-1}$.

We performed $^{27}$Al NMR spectroscopy and found Al species (23, 24) consistent with chloroaluminate anions and a urea-coordinated cation in the electrolytes (Fig.4 c,d).

Figure 4 c,d compare $^{27}Al$ NMR spectra of the $AlCl_3$-urea ILA to the $AlCl_3$-EMIC IL at the corresponding molar ratios. The spectrum of the $AlCl_3$/EMIC = 1.0 electrolyte showed a single peak corresponding to $AlCl_4^-$ (δ = 101.8 ppm) anion (Fig. 4c). However, the spectrum of $AlCl_3$/urea = 1.0 electrolyte showed four resonances: 52.7 ppm ($[AlCl_3 \cdot (urea)_2]$), 71.8 ppm ($[AlCl_2 \cdot (urea)_2]^+$), 88.0 ppm ($[AlCl_3 \cdot (urea)]$), and 101.5 ppm ($AlCl_4^-$)—assignments based on the work of Coleman et al. (22). The resonance at 52.7 ppm was broad and low intensity and is shown clearly in SI Fig. 2. In the $AlCl_3$/EMIC = 1.3 electrolyte, the system is totally ionic with $AlCl_4^-$ (δ = 101.8 ppm) and $Al_2Cl_7^-$ (δ = 96.7 ppm) being the dominant species at the 1.3 ratio. In the $AlCl_3$/urea = 1.3 electrolyte, the spectrum exhibited a much broader (likely due to chemical exchange (22)) feature than the $AlCl_3$/EMIC = 1.3, spanning the region corresponding to the anionic $AlCl_4^-$, $Al_2Cl_7^-$, and cationic species $[AlCl_2 \cdot (urea)_2]^+$, consistent with the existence of these ions in the electrolyte. Deconvolution of this broad resonance was performed to attempt to quantify the different species, but due to incurred difficulties the results were not considered for discussion.

**Conclusion**

A high efficiency battery that is stable over ~180 cycles and a variety of charge-discharge rates using an Al anode, graphite powder cathode, and cheap urea-$AlCl_3$ ionic liquid analogue electrolyte was successfully established. Intercalation/deintercalation of graphite during charging/discharging was confirmed by in-situ Raman experiments, and a stage 2 GIC was observed. Reversibility of the process was confirmed by recovery of the G-band at 1584 $cm^{-1}$ with no increase of the D-band intensity observed. Raman spectroscopy and $^{27}$Al NMR of the electrolyte suggested the presence of $AlCl_4^-$ , $[AlCl_2^-(urea)_n]^+$ and $Al_2Cl_7^-$ ionic species in the electrolyte.

The future prospects of the Al battery based on the urea-$AlCl_3$ electrolyte are promising and deserve further investigation. The high coulombic efficiency of the battery suggests long cycling capabilities, but this (ideally thousands of cycles) must be demonstrated. The earth abundance and low cost nature of the components of this battery make it a very attractive option for use on large scales, and its relatively low energy density (with respect to LIBs) is acceptable for non-mobile energy storage units. The rate capability of this battery is markedly less impressive than that of the EMIC-based battery system due to the higher viscosity and lower conductivity/ionicity of the electrolyte, but should have room for further improvement. While this work represents a satisfying step forward, exploration of numerous combinations of electrolytes and electrode materials remains wide open for further development of Al batteries to achieve ultra-high energy density/cost ratios.

**Note Added:** Bulk of this work was completed in 2015, and Stanford filed a patent on the Al-graphite-urea/$AlCl_3$ battery in December of 2015.

**Figures**

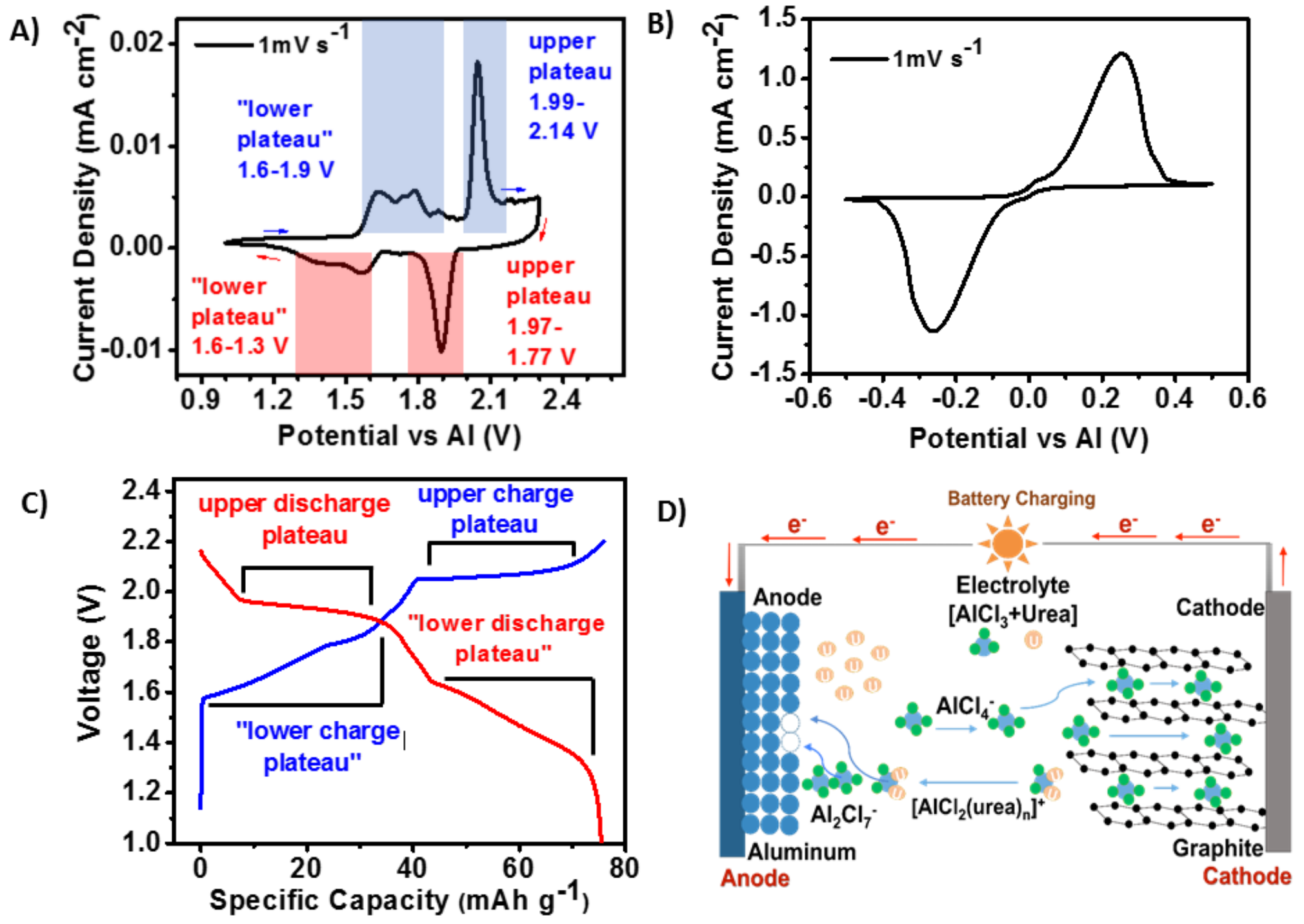

Figure 1. Cyclic voltammetry of 1.3 = $AlCl_3$/urea (by mol) at 1mV s-1 scan rate; **A)** graphite intercalation/deintercalation at graphite cathode, with corresponding major battery charge/discharge curve features indicated; **B)** aluminum deposition and stripping at Al anode **C)** Galvanostatic charge/discharge curve of $AlCl_3$ at 100mA $g^{-1}$. **D)** Schematic of battery discharge (Al dissolution at the anode, anion deintercalation at the cathode

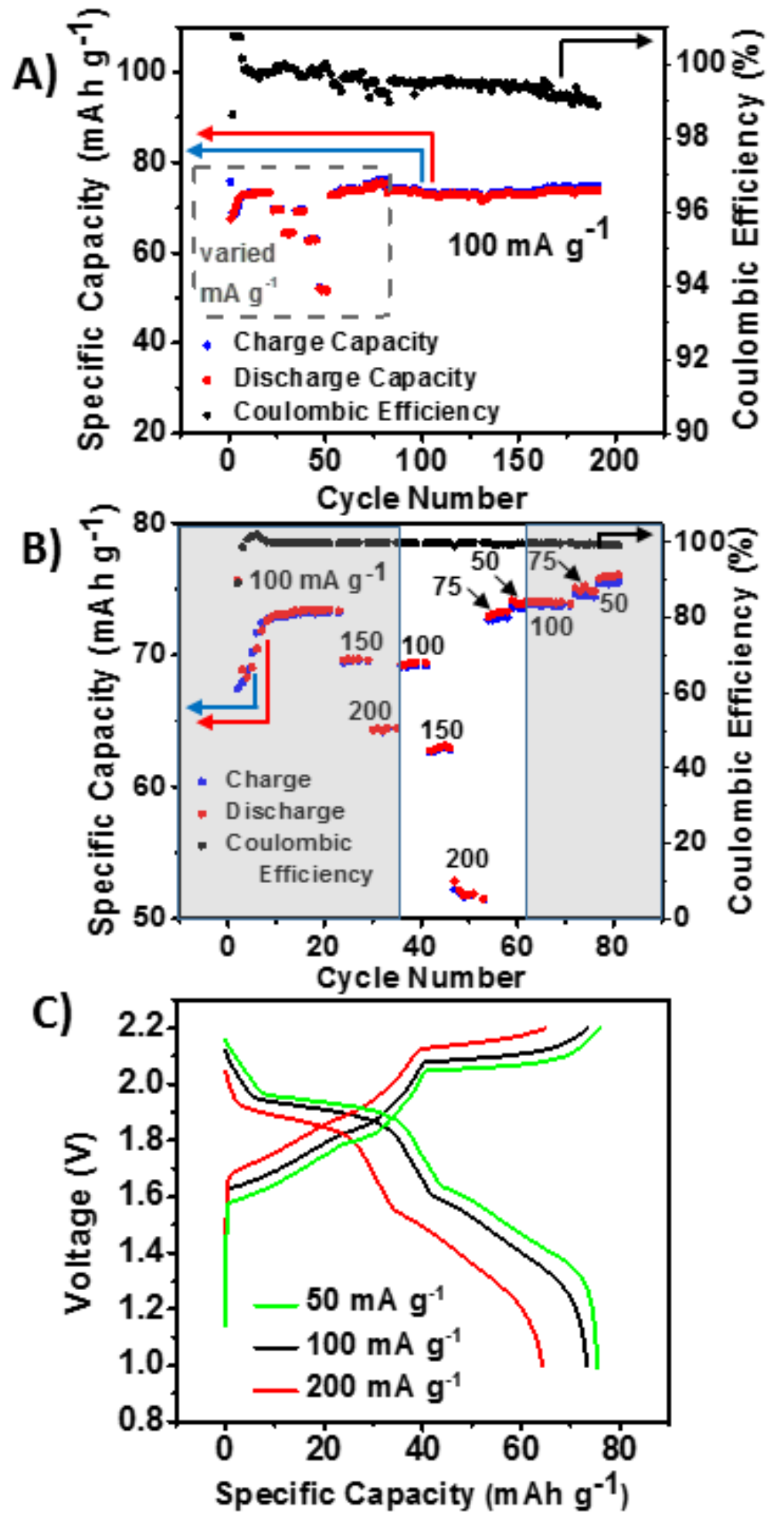

Figure 2. **A)** Stability test after charge-discharge rate variation for ~180 cycles (current density 100mA $g^{-1}$ and 2.2V/1V upper/lower cutoff); **B)** boxed region of A) (cycles 1-80) with varied charge/discharge rate. Region in gray depicts 2.2V upper cutoff, region in white depicts 2.15V upper cutoff). Lower cutoff is 1V for both. **C)** Galvanostatic charge-discharge curves for 50, 100, and 200 mA $g^{-1}$, 2.2V/1V upper/lower cutoff.

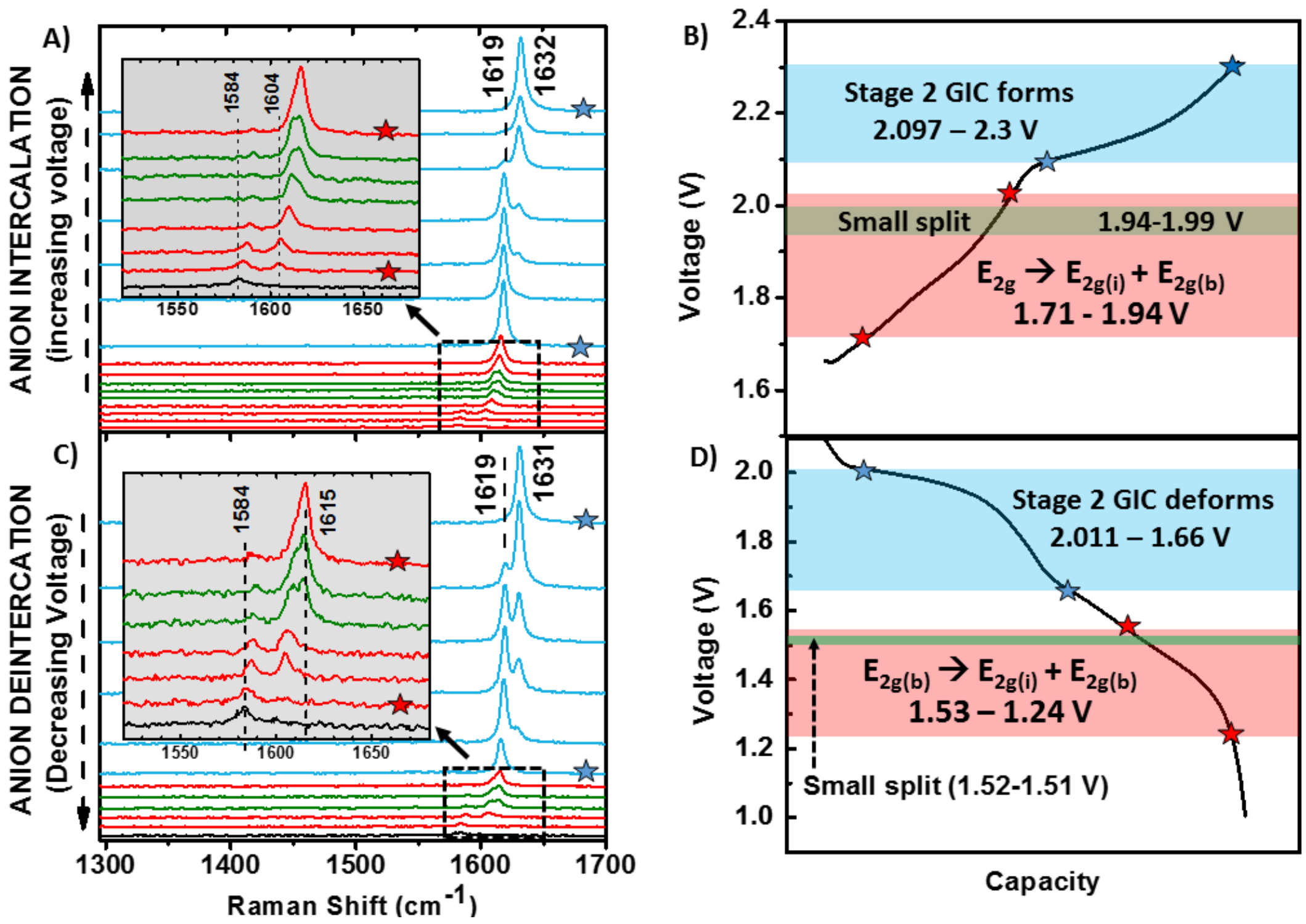

Figure 3. In-situ Raman spectra recorded during **A)** charge and **C)** discharge at 50 mA g$^{-1}$. Insets zoom in on lower voltage spectra corresponding to $E_{2g} \rightarrow E_{2g(i)} + E_{2g(b)}$ splitting (spectra in red, corresponding to red-shaded section of charge-discharge curves). The black spectrum in each corresponds to OCV = 1V, G-band = 1584 cm$^{-1}$. Spectra in blue (corresponding to upper plateau, shaded in blue on charge-discharge curves) represent stage 2 GIC formation/deformation. **B)** 50mA g $^{-1}$ Galvanostatic charging curve, color coordinated with Raman spectra in A). **D)** 50mA g $^{-1}$ Galvanostatic discharging curve, color coordinated with Raman spectra in C).

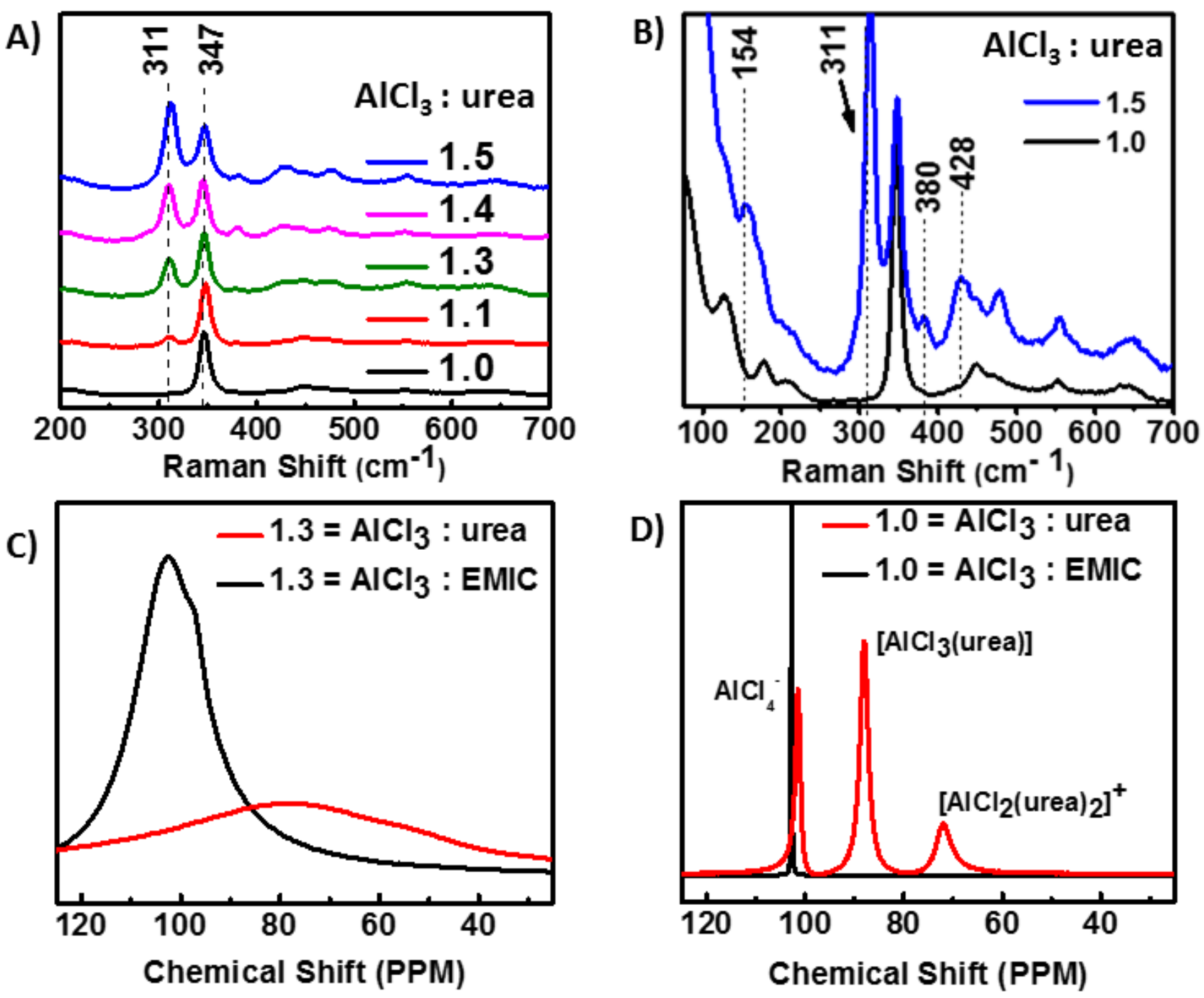

Figure 4. **A)** Raman spectra of 1.0, 1.3, 1.4, 1.5 = $AlCl_3$/urea electrolytes, normalized to peak at 347 $cm^{-1}$ ($AlCl_4^-$). **B)** Zoom of A) to elucidate lower intensity modes of $Al_2Cl_7^-$ (154, 310, 380, 428 $cm^{-1}$), 1.3, 1.4 = $AlCl_3$/urea electrolyte spectra excluded for clarity. $^{27}$Al NMR Spectra for **C)** $AlCl_3$/urea = 1.3 vs. $AlCl_3$/EMIC = 1.3 and **D)** $AlCl_3$/urea = 1.0 vs. $AlCl_3$/EMIC = 1.0.

**Materials and Methods**

SI Materials and Methods explains in detail the materials used and methods followed in this report.

**ACKNOWLEDGMENTS.** H. D. acknowledges support from the US Department of Energy DOE DE-SC0016165. B. H. acknowledges support from the Global Networking Talent 3.0 plan (NTUST 104DI005) from the Ministry of Education of Taiwan. M.-C.L. thanks the support from the Taishan Scholar Project for Young Scholars of Shandong Province of China.

**References**

1. Yang Z, *et al* (2011) Electrochemical energy storage for green grid. *Chem Rev* 111(5): 3577-3613.

2. Hogg JM, Coleman F, Ferrer-Ugalde A, Atkins MP & Swadzba-Kwasny M (2015) Liquid coordination complexes: A new class of lewis acids as safer alternatives to $BF_3$ in synthesis of polyalphaolefins. *Green Chem* 17(3): 1831-1841.

3. Abood HMA, Abbott AP, Ballantyne AD & Ryder KS (2011) Do all ionic liquids need organic cations? characterisation of $[AlCl_2 \cdot Amide]^+$ $AlCl_4^-$ and comparison with imidazolium based systems. *Chemical Communications* 47(12): 3523-3525.

4. Fang Y, *et al* (2015) An $AlCl_3$ based ionic liquid with a neutral substituted pyridine ligand for electrochemical deposition of aluminum. *Electrochim Acta* 160: 82-88.

5. Fang Y, Jiang X, Sun X & Dai S (2015) New ionic liquids based on the complexation of dipropyl sulfide and $AlCl_3$ for electrodeposition of aluminum. *Chemical Communications* 51(68): 13286-13289.

6. Pulletikurthi G, Boedecker B, Borodin A, Weidenfeller B & Endres F (2015) Electrodeposition of al from a 1-butylpyrrolidine-$AlCl_3$ ionic liquid. *Progress in Natural Science-Materials International* 25(6): 603-611.

7. Lin M, *et al* (2015) An ultrafast rechargeable aluminium-ion battery. *Nature* 520(7547): 324-328.

8. Wu Y, *et al* (2016) 3D graphitic foams derived from chloroaluminate anion intercalation for ultrafast aluminum-ion battery. *Adv Mater* 28(41): 9218-9222.

9. WILKES J, LEVISKY J, WILSON R & HUSSEY C (1982) Dialkylimidazolium chloroaluminate melts - a new class of room-temperature ionic liquids for electrochemistry, spectroscopy, and synthesis. *Inorg Chem* 21(3): 1263-1264.

10. Jung SC, Kang Y, Yoo D, Choi JW & Han Y (2016) Flexible few-layered graphene for the ultrafast rechargeable aluminum-ion battery. *Journal of Physical Chemistry C* 120(25): 13384-13389.

11. Jayaprakash N, Das SK & Archer LA (2011) The rechargeable aluminum-ion battery. *Chemical Communications* 47(47): 12610-12612.

12. Sun X*, et al* (2016) Polymer gel electrolytes for application in aluminum deposition and rechargeable aluminum ion batteries. *Chemical Communications* 52(2): 292-295.

13. Wang W*, et al* (2013) A new cathode material for super-valent battery based on aluminium ion intercalation and deintercalation. *Scientific Reports* 3: 3383.

14. Li J, Ma C, Chi M, Liang C & Dudney NJ (2015) Solid electrolyte: The key for high-voltage lithium batteries. *Advanced Energy Materials* 5(4): 1401408.

15. Smith AJ, Burns JC, Trussler S & Dahn JR (2010) Precision measurements of the coulombic efficiency of lithium-ion batteries and of electrode materials for lithium-ion batteries. *J Electrochem Soc* 157(2): A196-A202.

16. Balabajew M*, et al* (2016) In-situ raman study of the intercalation of bis(trifluoromethylsulfonyl) imid ions into graphite inside a dual-ion cell. *Electrochim Acta* 211: 679-688.

17. NEMANICH R, SOLIN S & GUERARD D (1977) Raman-scattering from intercalated donor compounds of graphite. *Physical Review B* 16(6): 2665-2672.

18. GALE R, GILBERT B & OSTERYOUNG R (1978) Raman-spectra of molten aluminum-chloride - 1-butylpyridinium chloride systems at ambient-temperatures. *Inorg Chem* 17(10): 2728-2729.

19. TAKAHASHI S, CURTISS L, GOSZTOLA D, KOURA N & SABOUNGI M (1995) Molecular-orbital calculations and raman measurements for 1-ethyl-3-methylimidazolium chloroaluminates. *Inorg Chem* 34(11): 2990-2993.

20. Gilbert B, Olivier-Bourbigou H & Favre F (2007) Chloroaluminate ionic liquids: From their structural properties to their applications in process intensification. *Oil & Gas Science and Technology-Revue D Ifp Energies Nouvelles* 62(6): 745-759.

21. Coleman F, Srinivasan G & Swadzba-Kwasny M (2013) Liquid coordination complexes formed by the heterolytic cleavage of metal halides. *Angewandte Chemie-International Edition* 52(48): 12582-12586.

22. Hu P*, et al* (2016) Structural and spectroscopic characterizations of amide-AlCl3-based ionic liquid analogues. *Inorg Chem* 55(5): 2374-2380.

23. DEROUAULT J, GRANGER P & FOREL M (1977) Spectroscopic investigation of aluminum trihalide-tetrahydrofuran complexes .2. solutions of aluminum-chloride or bromide in tetrahydrofuran and in tetrahydrofuran-dichloromethane. *Inorg Chem* 16(12): 3214-3218.

24. GRAY J & MACIEL G (1981) Al-27 nuclear magnetic-resonance study of the room-temperature melt $AlCl_3$-normal-butylpyridinium chloride. *J Am Chem Soc* 103(24): 7147-7151.

**Supporting Information**

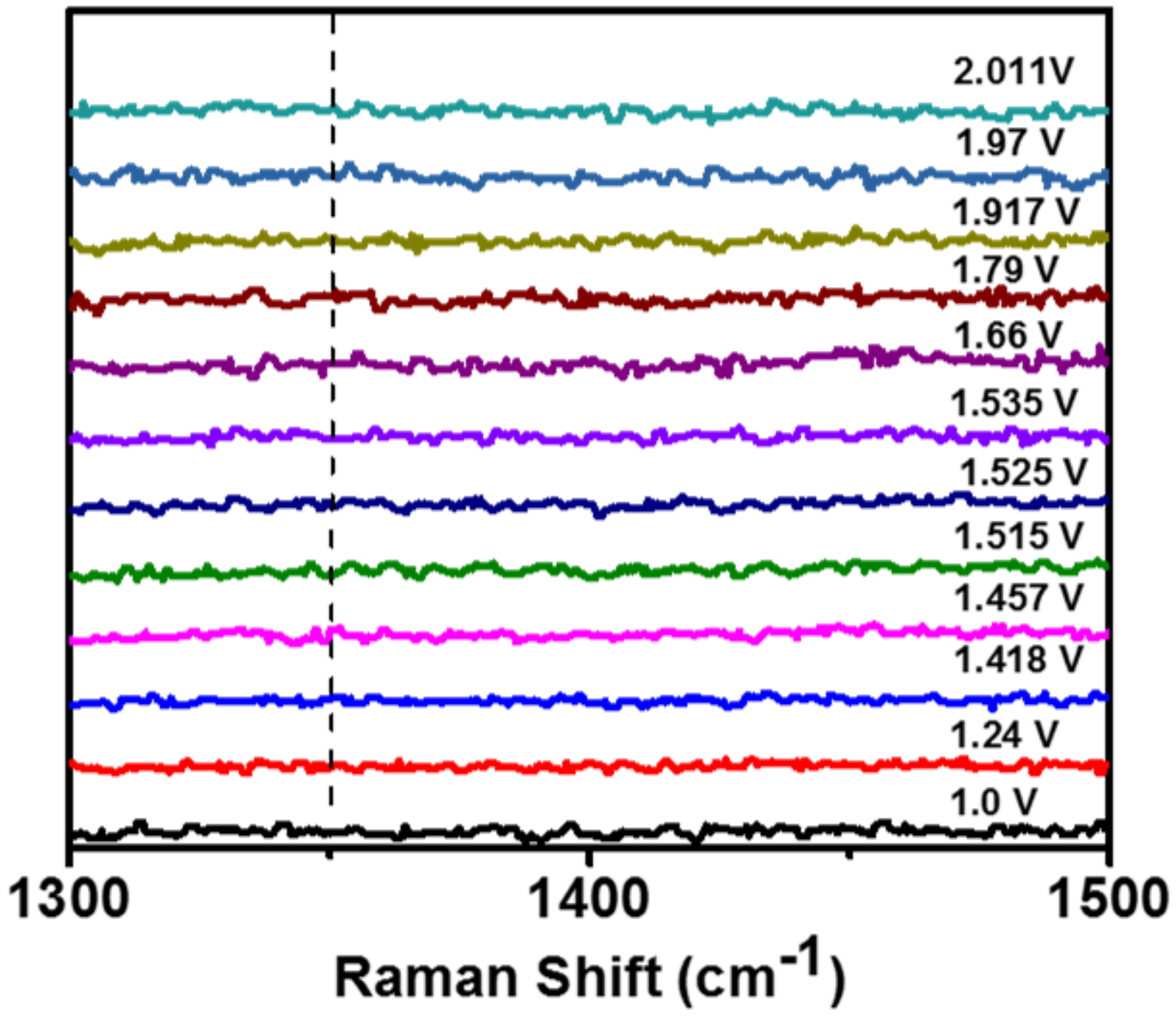

SI Figure 1: In-situ Raman spectroscopy during discharge (50mA $g^{-1}$), D-band region. No D-band is detectable before or after discharge. This was the cell's 81st cycle.

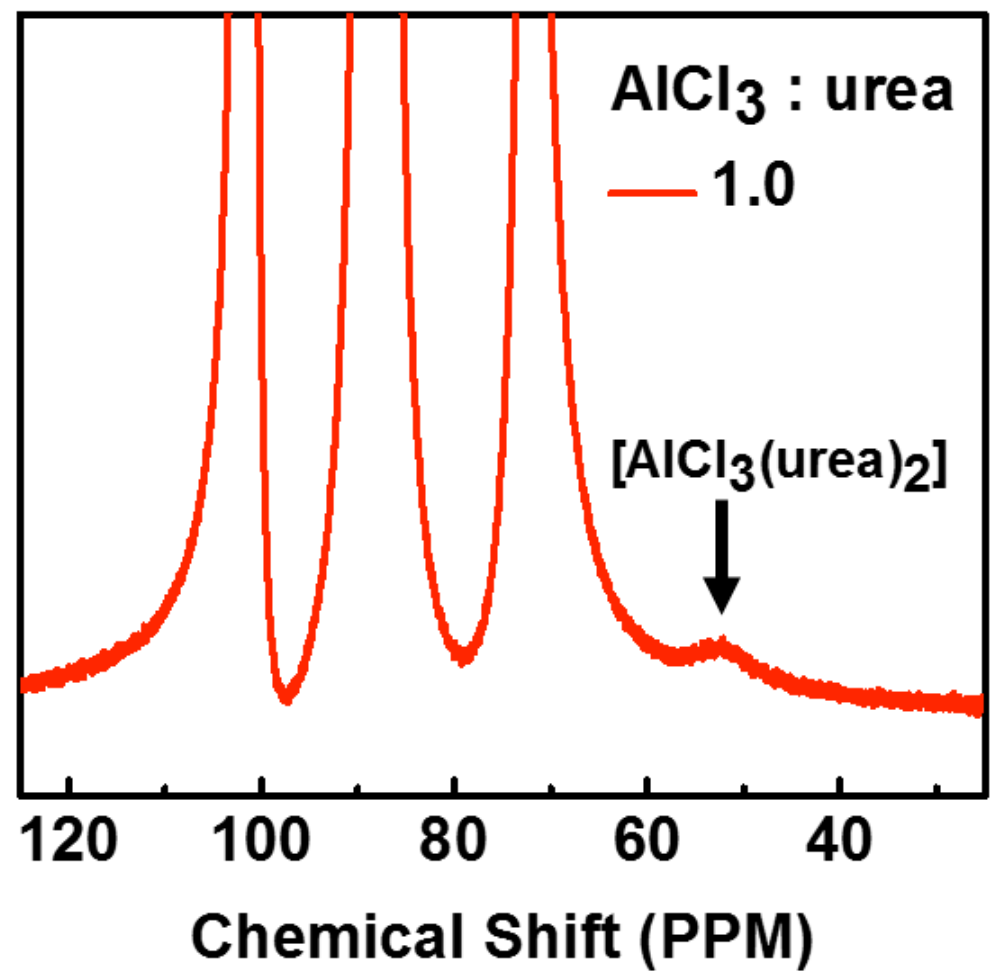

SI Figure 2. $^{27}Al$ spectrum of $AlCl_3$/urea = 1.0. Zoom at 52.7 ppm ([$AlCl_3 \cdot (urea)_2$]) to show broad, low intensity resonance.

## Materials and Methods

### Preparation of $AlCl_3$-urea ILA (1.3 = $AlCl_3$-urea by mol) battery electrolyte:

2g of urea (VWR, 99.9% ultrapure) and 5.77g anhydrous $AlCl_3$ (Fluka, ≥99%, crystallized) were added in small portions with constant magnetic stirring.

### NMR measurements:

$^{27}Al$ NMR spectra were recorded using a UI300 spectrometer at 300 MHz (64 scans, acquisition time 0.5s) referenced to 1.1M $Al(NO_3)_3$ in $D_2O$. All spectra were recorded for neat samples without a lock, and the temperature was calibrated against methanol (± $1^oC$).

### Raman measurements (ILAs):

Neat samples of 1.0, 1.1, 1.3, 1.4, and 1.5 = $AlCl_3$ /urea by mol prepared as described above and spectra were acquired using an $Ar^+$ laser (532 nm) with 0.8 $cm^{-1}$.

### Electrochemical Measurements:

All cells were made in aluminum laminated pouch cell cases (MTI, EQ-alf-100-210). High purity Al foil, 3mm Nickel tab (MTI, EQ-PLiB-NTA3), graphite powder, and polymer binder were used.

*Batteries*

Assembled cells were baked at $80^oC$ under vacuum overnight. For construction of the pouch cell, A Ni tab was used as a current collector, which could then be heat sealed. 1.5g of 1.3 = $AlCl_3$-urea by mol electrolyte were inserted inside the glovebox. Galvanostatic charge/discharge measurements were performed outside the glovebox.

*Cyclic Voltammetry (CV)*

Cyclic voltammetry measurements were performed on a potentiostat/galvanostat model CHI 760D (CH Instruments). Aluminum electrodes were washed with acetone and gently scrubbed with a kim wipe before cell assembly. One layer of separator material was used. All CVs were performed at a scan rate of 1mV $s^{-1}$.

### In-situ Raman measurements:

In-situ Raman cells were constructed as previously described (7). Galvanostatic charge/discharge was performed for ~80 cycles at 100mA $g^{-1}$ to ensure regular battery behavior, then at 50mA $g^{-1}$ (~0.66C) while spectra were recorded (2s acquisition time, 5 scans) for every 0.01V change. Selected spectra were chosen for Figure 3.